\documentclass{iaus}
\pdfoutput=1

\usepackage{graphicx}
\newcommand\nH{n_\mathrm{H}}

\newcommand\la{\lesssim}
\newcommand\percc{\mathrm{cm}^{-3}}
\newcommand\persqcm{\mathrm{cm}^{-2}}

\title[Collisional excitation of OH(6049\,MHz) masers] 
{Collisional excitation of OH(6049\,MHz) masers in supernova remnant
-- molecular cloud interactions}

\author[Wardle]   
{Mark Wardle}

\affiliation{Department of Physics, Macquarie University,
Sydney, NSW 2109, Australia \break email: wardle@physics.mq.edu.au}

\pubyear{2007}
\volume{242}  
\pagerange{119--126}
\date{?? and in revised form ??}
\jname{Astrophysical Masers and their Environments IAU Symposium}
\editors{J.M.\ Chapman \& W.A.\ Baan, eds.}
\begin{document}

\maketitle

\begin{abstract}
OH (1720 MHz) masers serve as indicators of SNR -- molecular cloud
interaction sites.
These masers are collisionally excited in warm (50-100\,K) shocked gas
with densities of order $10^5\percc$ when the OH column density is
in the range $10^{16}$--$10^{17}\persqcm$.  Here I present
excitation calculations which show that when the OH column density
exceeds $10^{17}\persqcm$ at similar densities and temperatures,
the inversion of the 1720 MHz line switches off and instead the 6049
MHz transition in the first excited rotational state of OH becomes
inverted.  This line may serve as a complementary signal of warm,
shocked gas when the OH column density is large.
\keywords{supernova remnants, molecular processes, radiation 
mechanisms: nonthermal, radio lines: ISM}
\end{abstract}


The OH excitation calculations discussed here include the 32 lowest
energy levels, with transition wavelengths and A values from
Detombes et al.\ (1977) and Brown et al.\ (1982), and rates for
collisional deexcitation by H$_2$ kindly provided by Alison Offer
(private communication).  Following Lockett \& Elitzur (1989) and
Lockett, Gauthier \& Elitzur (1999), I adopt a uniform slab model for
the masing medium.  Radiative transfer is approximated using escape
probabilities based on the mean optical depth, including the effects
of line overlap.  The slab is parametrised by density $\nH$,
temperature $T$, column $N_\mathrm{OH}$, velocity width, and a
radiation field contributed by the CMB and warm dust.

\begin{figure}\centering
\includegraphics[scale=0.4]{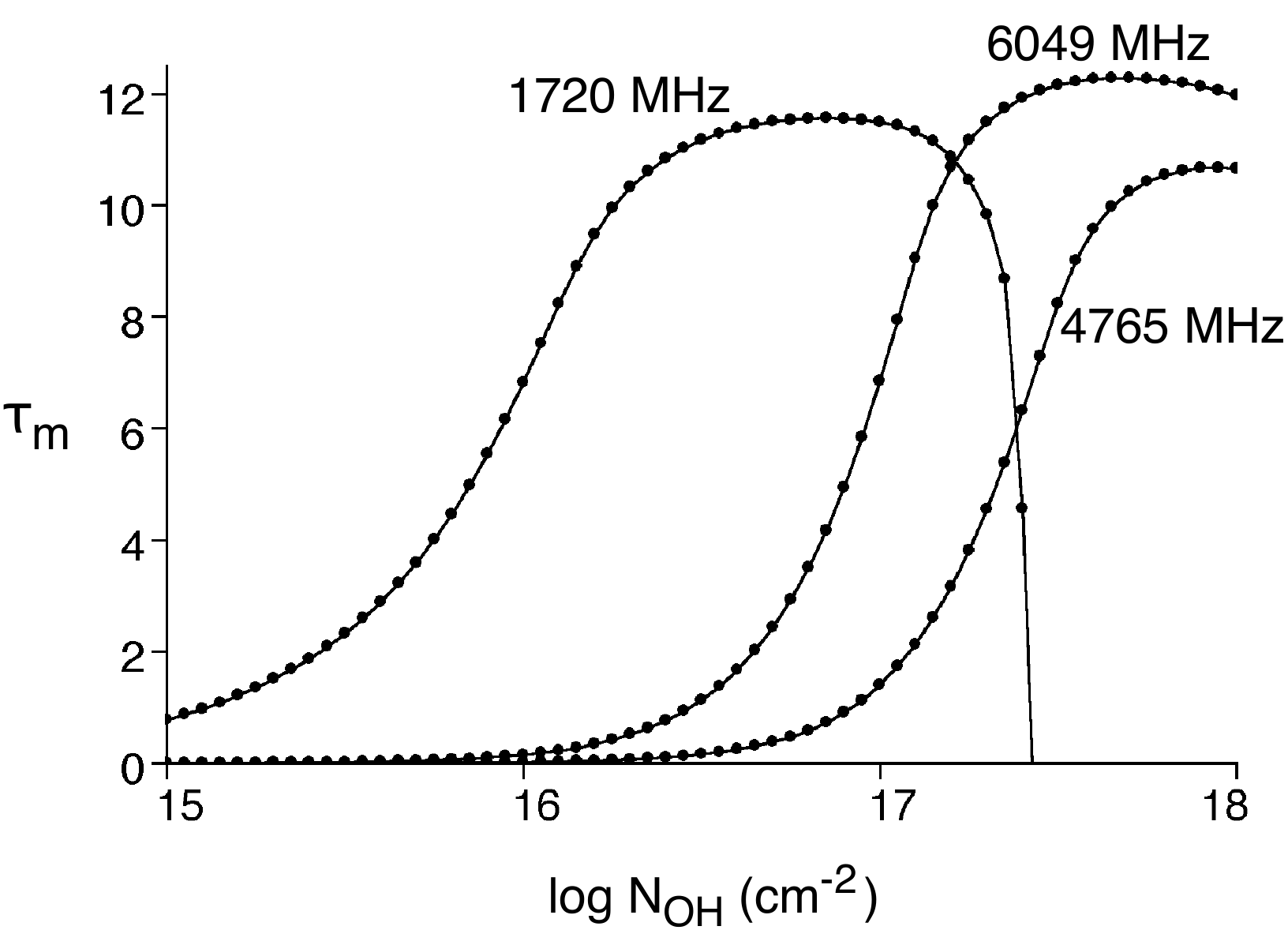}
  \caption{ Maser optical depth, $\tau_m$ ($I_\nu\propto
\exp(\tau_m)$)  of the 1720 MHz OH line
and its analogues at 6049 and 4765\,MHz, as a function of OH column density for T = 50\,K and
$\nH = 10^5 \percc$. }\label{fig:tau}
\end{figure}

\begin{figure}
    \centering
\includegraphics[scale=0.5]{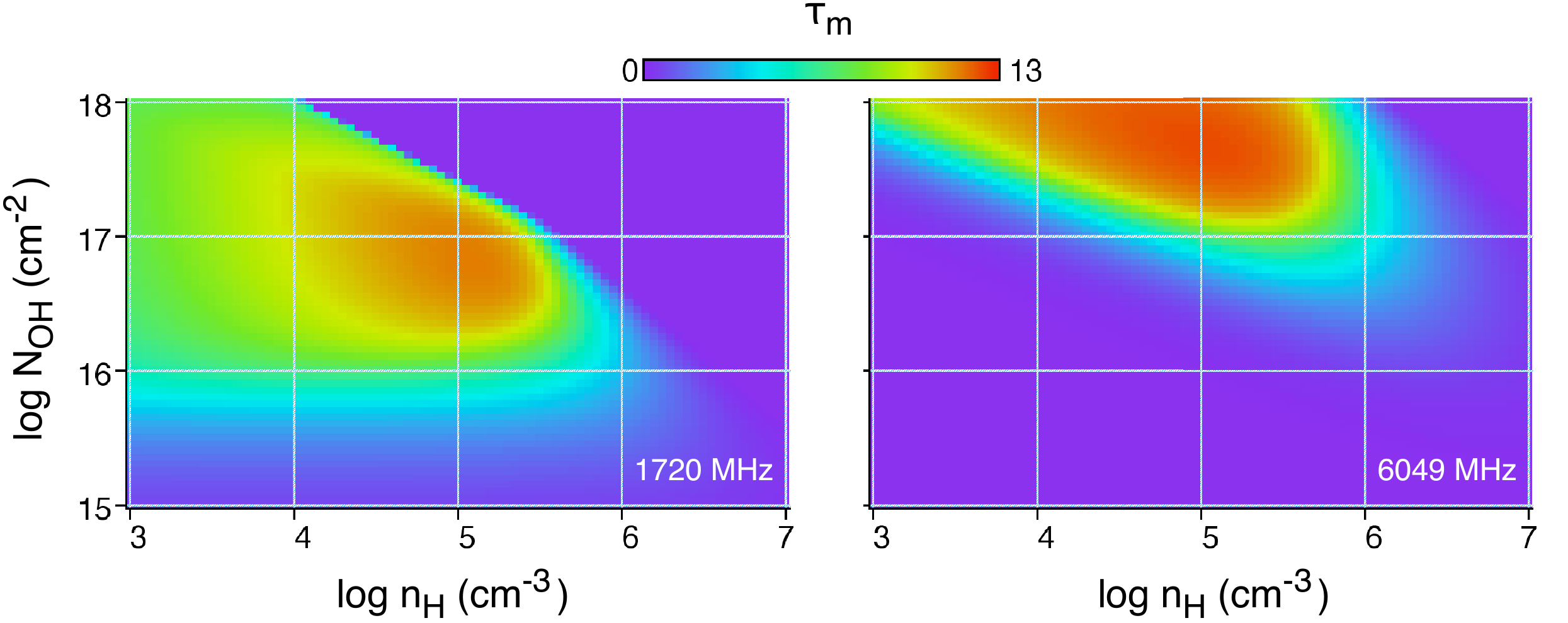}
  \caption{Maser optical depth as a function of $\nH$
and $N_\mathrm{OH}$ for T = 50\,K for the 1720 MHz (left) and 6049
MHz (right) lines of OH.}\label{fig:contours}
\end{figure}

Figs.\ 1 and 2 show results for $T=50$\,K, assuming small
velocity gradients within the slab, and no far-infrared
radiation from dust.  These conditions are broadly consistent with
warm gas that is cooling off behind a shock wave driven into
a molecular cloud by an adjacent supernova remnant.  Fig.\
\ref{fig:tau} shows the maser optical depth through the slab
for the 1720 MHz satellite line in the ground rotational state and its
analogues in the first (6049 MHz) and second (4765 MHz) excited
rotational states.  At low OH column densities the 1720 MHz line is
inverted, peaking in the range $10^{16}$--$10^{17}\persqcm$.  At about
$10^{17}\persqcm$ the optical depth of the slab to the 1720 MHz line
supresses the inversion.  Meanwhile the inversion in the 6049\,MHz
line grows, peaking at $N_\mathrm{OH}\approx 3\times 10^{17}\persqcm$ 
(similar results were found by Pavlakis \& Kylafis 2000).
The 4765\,MHz line peaks at $N_\mathrm{OH}\sim 10^{18}\persqcm$.
  The effect
of varying $\nH$ and $N_\mathrm{OH}$ is explored in Fig.\
\ref{fig:contours}.  As expected, masing in the 1720\,MHz line is
strongest for $\nH\sim 10^5\percc$ and $N_\mathrm{OH}\sim
10^{16.5}\persqcm$.  The inversion of the 6049\,MHz line requires
column densities in excess of $10^{17}\persqcm$, while inversion at
4765\,MHz (not shown) requires an OH column a few times higher still.

OH (1720 MHz) masers are a signature of SNR -- molecular cloud
interactions (Frail et al.\ 1994; Wardle \& Yusef-Zadeh 2002).
Masing at 6049\,MHz may take over this role when the OH column density
exceeds $10^{17}\persqcm$, too high for 1720 MHz masers to exist.
Note, however, that the OH column density typically produced in
SNR--cloud interactions is uncertain.  Existing models of OH
production rely on UV or X-ray dissociation of the water produced in
molecular shocks, and yield column densities $\la 10^{16}\persqcm$
(Lockett et al.\ 1999; Wardle 1999).  As the dissociation rate of OH
is approximately half that of H$_2$O, these models predict
OH/H$_2\mathrm{O}\la 2$, and typically much less.  This conflicts with
recent absorption measurements in IC 443, which find that
$N_\mathrm{H_2O}\sim 10^{14-15}\persqcm$ (Snell et al.\ 2005) and
$N_\mathrm{OH}\sim 10^{16-17}\persqcm$ (Hewitt et al.\ 2006).
Preliminary analysis of a survey for 6049\,MHz masers towards southern
SNRs has not yielded detections to date (see McDonnell, Vaughan \&
Wardle, elsewhere in these proceedings).

\end{document}